\documentclass{aa}
\usepackage{graphicx}
\usepackage{supertabular}
\begin{document}
   \title{Early decline spectra of Nova SMC 2001 and Nova LMC 2002}


   \author{E. Mason,
          \inst{1}
	   M. Della Valle,
	  \inst{2} 
	   R. Gilmozzi,
	  \inst{1} 
	  G. Lo Curto,
	  \inst{1} \and
	   R. E. Williams
	  \inst{3} 
          }

   \offprints{E. Mason}

   \institute{ESO, Alonso de Cordova 3107, Casillas 19001, Vitacura, Santiago, Chile\\
              \email{emason@eso.org}
	\and
	Osservatorio Astronomico di Arcetri, Largo E. Fermi 5, Firenze, Italy
	\and
	Space Telescope Science Institute, 3700 San Martin Drive, Baltimore, MD, USA
             }

   \date{Received ?}

   \abstract{We report results on the spectroscopic follow-up of Nova SMC
2001 and Nova LMC 2002 carried out at La Silla. The analysis of the
spectroscopic evolution shows that these objects belong to the {\sl Fe
II} class, according to the Cerro Tololo scheme. From the line fluxes
and the expansion velocities, we have derived an approximate
mass for the ejected shells of 2$\div$3$\times 10^{-4} M_\odot$. The
filling factor measurements ($\varepsilon\sim 10^{-4}\div 10^{-1}$)
suggest a clumpy structure for the ejecta.  }

   
\authorrunning{Mason et al.}

   \maketitle
%

\section{Introduction}

Spectroscopic observations of extra-galactic novae are relatively
rare. In this paper we report the spectroscopic evolution of two
novae that recently occurred in the Magellanic Clouds: Nova SMC 2001 and
Nova LMC 2002. These data are doubly interesting: they provide
important information on the nova event itself and help
to characterize the properties of nova populations in galaxies that have
different metallicity content, luminosity class and Hubble
type.
 The existence of systematic
differences among extra-galactic novae
has been recognized on the basis of the observed nova counts
 (Duerbeck 1990), the distribution of the photometric speed classes
(Della Valle et al. 1992), the frequency of occurrence of nova events in
different Hubble type galaxies (Della Valle et al. 1994, Williams and
Shafter 2004) and the
spectroscopic evolution (Della Valle \& Livio
1998).

Nova SMC 2001 was discovered by Liller (2001) on Oct 21.09 UT at a
magnitude V=12.21. The first spectra were taken on Oct 26.15 UT and 30.25 UT by
Bosch et al. (2001) and Della Valle et al. (2001).
They reported the detection of strong Balmer
and low ionization emission lines (e.g., FeII, OI,
and NaI), flanked by P-Cyg profiles which indicated an expansion
velocity of 950 km/sec from the P-Cyg absorption of the FeII and NaI lines,
while velocities of 1400 and 1800 km/sec were derived from the P-Cyg
absorption of the Balmer lines H$\alpha$ and H$\beta$, respectively.
Jensen et al.(2001) reported the detection
 of CII in a spectrum obtained on Nov 2.12 UT
and pointed out that the nova strongly
resembles Nova LMC 1988 No.1 two weeks after maximum. This may suggests that
nova SMC 2001 reached maximum light around October 15.

Nova LMC 2002 was discovered by Liller (2002) on March 3 when
the object was close to maximum brightness. The nova was also
detected in two pre-discovery images (Liller 2002) obtained on
February 21 and 27 (m$_{pg}\sim$15 and m$_{pg}$=12.5, respectively).
In the first low resolution spectrum, taken on March 4, Liller (2002) reports
the presence of strong H$\alpha$ and H$\beta$ emission.
After Liller's discovery there are only few photometric
observations by Kilmartin \& Gilmore (2002) and Gilmore (2002).
We plot in Fig.~1 the visual photometry in the IAU Circulars (No. 7841,
7847 and 7853) and the V magnitudes estimated from our spectra
(see Sec.~3 and Table~4).

   \begin{figure} \centering
   \rotatebox{-90}{\includegraphics[width=6.5cm,]{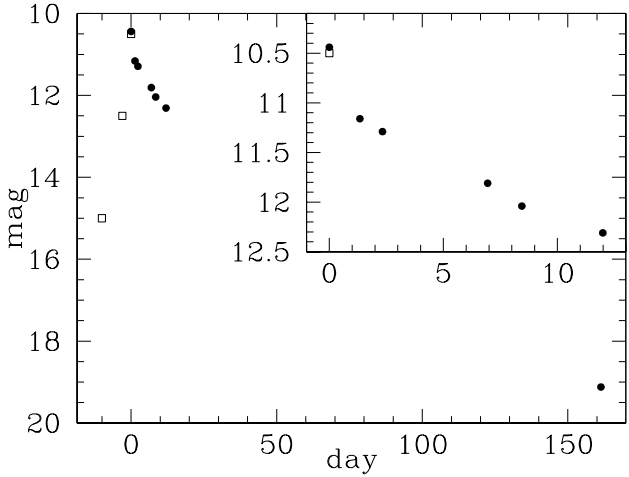}}
   \caption{Nova LMC 2002 light curve. Day 0 is the day of the
   observed maximum. The upper right box is a zoomed-in view of the
   light curve. Squares in the rising part of the curve are
   photographic magnitudes, filled circles are V band magnitudes.}
   \label{f1}
   \end{figure}

We present our spectroscopic observations in Sec.~2. The data and their analysis are reported in Sec.~3.  The summary and the conclusion follow in Sec.~4.

   \begin{table*}
         \label{log}
 \begin{center}
\scriptsize
\caption{The log of observations of Nova SMC 2001.}
\begin{tabular}{ccccccc}
 & & & & & & \\
\hline
Obs. date & UT start & Exp. time (sec) & Telescope+Instrument & mode & $\lambda$ range & \AA/pix \\
\hline
 & & & & &  \\
01/11/2001 & 01:12 & 1800 & 1.5+FEROS & (echelle)  & 3900-9200 & 1.0\\
06/12/2001 & 01:20 & 3722 & NTT+EMMI  & DIMD2, grating\#3  & 3300-3750 & 0.45\\
 & 02:27 & 2522 & & & 3500-4000 & 0.45\\
 & 03:15 & 2522 & & & 4000-4500 & 0.45\\
 & 04:02 & 2522 & & & 4500-5000 & 0.45 \\
 & 01:20 & 3600& & DIMD1, grating\#7 & 4800-6200 & 0.67\\
 & 02:27 & 2522& & & 6150-7500 & 0.65\\
 & 03:16 & 2400& & & 7450-8750 & 0.65\\
 & 04:02 & 2400& & & 8750-10050 & 0.64\\
17/02/2002 & 02:01 & 300 & NTT+EMMI & RILD, grism\#1  & 3600-10200 & 5.8\\
 & 01:03 & 1200 & & RILD, grism\#5 & 4000-6650 & 1.3 \\
 & 01:27 & 1800 & & RILD, grism\#6 & 6000-8350 & 1.2 \\
4/03/2002 & 01:27 & 600 & NTT+EMMI & RILD, grism\#1 & 3600-10200 & 5.8 \\
 & 00:16 & 1550 & & RILD, grism\#5 & 4000-6650 & 1.3\\
 & 00:46 & 2265 & & RILD, grism\#6 & 6000-8350 & 1.2\\
\hline
\end{tabular}
\end{center}
   \end{table*}

   \begin{table*}
         \label{log2}
 \begin{center}
\scriptsize
\caption{The log of observations of Nova LMC 2002.}
\begin{tabular}{ccccccc}
 & & & & & & \\
\hline
Obs. Date & UT start & Exp. time (sec) & Telescope+Instrument & mode & $\lambda$ range (\AA) & \AA/pix \\
\hline
 & & & & & & \\
09/03/2002 & 23:55,00:00,00:11 & 40+60$\times$2    & 3.6+EFOSC & grism\#9 & 4700-6750 & 2.0 \\
 & 23:59,00:08,00:19 & 20+35$\times$2 & & grism\#12 & 6000-10300 & 4.2 \\
 & 23:56,00:02,00:12 & 40+100$\times$2 & & grism\#14 & 3200-5100 & 2.0 \\
 & 23:57,00:04,00:14 & 40+200$\times$2 & & grism\#15 & 6900-8750 & 1.9 \\
14/03/2002 & 01:14,01:24 & 60+120 & 3.6+EFOSC & grism\#9 & 4700-6750 & 2.0\\
 & 01:22,01:37 & 35+70 & & grism\#12 & 6000-10300 & 4.2 \\
 & 01:15,01:26 & 100+200 & & grism\#14 & 3200-5100 & 2.0 \\
 & 01:26,01:30 & 200+400 & & grism\#15 & 6900-8750 & 1.9 \\
11/08/2002 & 10:15 & 900 & NTT+EMMI & RILD,grism\#3 & 3850-9100 & 2.8 \\
\hline
\end{tabular}
\end{center}
   \end{table*}

\section{Observations}

\begin{table}
\begin{center}
\scriptsize
\caption{The observed spectroscopic evolution of Nova SMC 2001 and
Nova LMC 2002 according to the Cerro Tololo classification. For Nova
SMC~2001 we write in brackets the spectroscopic phase reported in the
IAUC No.7744.}
\begin{tabular}{ll}
& \\
\hline
Object & Spec. ev. \\
\hline
& \\
Nova SMC 2001 & $(P_{fe}^o)$,$P_{na/n}^o$,$P_n^o$, $A_o$ \\
Nova LMC 2002 & $P_{fe}$, $A_o$ \\
\hline
\end{tabular}
\end{center}
\end{table}

Spectroscopic observations for Nova SMC 2001 and Nova LMC 2002 were
taken in service mode
during the target of opportunity campaign at the 1.5m+FEROS,
3.6m+EFOSC, and NTT+EMMI, at La Silla Observatory.  Details of the log of
observations are in Table~1 and 2 for Nova SMC 2001 and Nova LMC 2002,
respectively.
Spectrophotometric standard stars were taken every night except November 1.
Screen flats, bias, and wavelength calibration frames
were secured on the day following each observing run. The spectra were
reduced using standard IRAF routines.

\begin{table}
\begin{center}
\scriptsize
\caption{ The instrumental magnitudes derived by convolution of the
observed spectra with the standard V-Johnson filter.}
\begin{tabular}{cc}
\hline
date & V \\
\hline
\multicolumn{2}{c}{Nova SMC 2001} \\
\hline
06/12/01 & 15.3 \\
17/02/02 & 16.4 \\
04/03/02 & 15.4 \\
\hline
& \\
\hline
\multicolumn{2}{c}{Nova LMC 2002} \\
\hline
09/03/02 & 11.8 \\
14/03/02 & 12.3 \\
11/08/02 & 19.1 \\
\hline
\end{tabular}
\end{center}
\end{table}

\begin{table*}
\begin{center}
\scriptsize

\caption{Nova SMC 2001 emission line fluxes. The absolute fluxes for
November 2001 have been derived from the relative fluxes
after assuming t2$\sim 20$ days. Flux measurements are
affected by errors of the order of $\sim50\%$ (col. 2) and $\sim 30\%$
(cols. 3, 4, 5).}

\begin{tabular}{p{5cm}p{2.5cm}p{2.5cm}p{2.5cm}p{2.5cm}}
 & & & &  \\
\hline
Line & 01-11-2001 & 06-12-2001 & 17/18-02-2002 & 4/5-03-2002 \\
\hline
& & &  & \\
H18(4)  3692  & --- & 7.54E-15 & --- & ---\\
H17(3) 3697 & --- & 1.58E-14 & & \\
H13(3) 3734 & ---& 3.96E-14 & ---& ---\\
H12(2)  3750 & --- & 7.63E-14 & ---& ---\\
H11(2)  3771 & --- & 9.25E-14 & ---& ---\\
H10(2)  3798 & ---& 6.88E-14 &--- &--- \\
H$\eta$  3835 & ---& 1.16E-13 & ---& ---\\
H$\zeta$ 3889 & ---& 1.44E-13 & ---& ---\\
H$\epsilon$ 3970 & 1.58E-12 + CaII(1) & 1.54E-13 & 6.37E-14 & ---\\
HeI(18) 4026 & --- & --- & 1.10E-14 & --- \\
$[SII]$(1) 4068 & --- & --- & 3.05E-14 & 3.21E-14 \\
H$\delta$ 4102 & 4.83E-12  & 3.19E-13 & 1.44E-13 & 1.35E-13 \\
FeII(28) 4179 & 3.89E-12 & --- & --- & --- \\
HeII(3) 4200 & --- & --- & 1.61E-14 & 1.34E-14 \\
FeII(27) 4233 & 1.02E-12 & 1.16E-14 & --- & --- \\
CII(6) 4267  & 7.37E-13 & 5.10E-14 & 2.03E-14 & ---\\
H$\gamma$ 4341  & 4.71E-12 & 5.145E-13 & 1.29E-13 & 1.218E-13 \\
$[OIII]$(2) 4363 &  --- &  -- & 4.77E-13 & 3.91E-13 \\
FeII(27) 4417 & 7.78E-13 & 2.47E-14 & --- & --- \\
$[FeII]$(7) 4452 & 1.64E-13 & 2.33E-14 & --- & --- \\
HeI(14) 4471 & --- & --- & 1.47E-14 &  \\
$[FeII]$(7) 4475  & 2.26E-13 & 3.62E-14 & --- & --- \\
MgII(4) 4481 & 1.64E-12 & --- & --- & --- \\
FeII(37) 4515/20 & 1.64E-13 & --- & --- & --- \\
NIII(3) 4511/15/24 & --- & --- & 2.77 E-14 & 2.65E-14 \\
FeII(38) 4523/40 & 4.71E-13 & --- & --- & --- \\
FeII(38) 4550 + (37) 4556 & 2.05E-13 & --- & --- & --- \\
FeII(38) 4584 & 3.48E-13 & --- & --- & --- \\
NII(5) 4607/14 & --- & --- & 3.03E-14 & 3.01E-14 \\
NIII(2) 4641/2 + CIII(1) 4650 & --- & 9.57E-13 & 1.40E-13 &  1.36E-13 \\
OII(1) 4651/61 & 1.47E-12 & --- & --- & --- \\
HeII(1) 4686 &--- & --- & 5.13E-13 & 4.81E-14 \\
NII(20) 4803/10 & --- & --- & 7.43E-15 & 5.43E-15 \\
H$\beta$ 4861 & 1.29E-11 & 8.13E-13 & 2.25E-13 & 2.37E-13 \\
HeI(48) 4922 & --- & 1.17E-13 &  1.044E-13 & --- \\
FeII(42) 4924 & 3.89E-12 & --- & --- & --- \\
$[OIII]1$ 4959 & --- & 5.82E-14 & 2.70E-13 & 3.67E-13  \\
$[OIII]1$ 5007 & --- & 4.05E-13 & 7.78E-13 & 1.06E-12 \\
HeI(4) 5016 & --- & --- & 2.00E-14b & 1.8E-14b \\
FeII(42) 5018 & 4.91E-12 & --- & --- & --- \\
FeII(42) 5169 & 4.10E-12 & --- & --- & --- \\
NII(66) 5176/80 & --- &  4.19E-14 & 1.33E-14 & 9.80E-15 \\
FeII(49) 5198 & 1.13E-12 & --- & --- & --- \\
FeII(49) 5235 & 7.57E-13 & --- & --- & --- \\
FeII(49) 5276 & 1.27E-12 & 2.96E-14 & --- & --- \\
FeII(49) 5317 & 2.25E-12 & ---& --- & ---\\
FeII(48) 5363 & 1.04E-12  & --- & 1.369E-15 & 1.569E-15 \\
HeII(2) 5412 & --- & --- & 3.53E-15 & 3.54E-15 \\
NII(29) 5463/80/96 & --- & 2.23E-14 & 4.10E-15 & 3.56E-15 \\
FeII(55) 5535 & 9.62E-13 & --- & --- & --- \\
$[OI]$(3) 5577 & 3.89E-12 & 3.55E-14  & 2.54E-15 &  2.81E-15 \\
NII(3) 5666/80 & 1.02E-11 &  1.39E-13 & 2.62E-14 & 2.66E-14 \\
\end{tabular}
\end{center}
\end{table*}

\begin{table*}
\begin{center}
\setcounter{table}{4}
\scriptsize
\caption{continued}
\begin{tabular}{p{5cm}p{3cm}p{3cm}p{3cm}p{3cm}}
$[NII]$(3) 5755 & 1.15E-11 & 4.04E-13 & 1.53E-13 & 2.05E-13 \\
CIV(1) 5802/12 &  ---   & 1.16E-13 & 5.36E-15 & 4.99E-15 \\
HeI(11) 5876 	& ---	& 1.25E-13 & 2.69E-14 & 2.75E-14 \\
NaI(1) 5896 & 1.02E-11 & ---& ---&--- \\
NII(28) 5940/42 & 8.80E-12  & 3.80E-14 & 8.86E-15 & 6.84E-15 \\
FeII(46) 5991 + FeII(24) 6021 & 5.94E-12 & ---& ---& ---\\
CaI(3) 6103   & 2.46E-12 & 3.42E-14  &--- & --- \\
FeII(74) 6149 &  6.35E-12 & 2.38E-14 & 4.11E-15  & 3.65E-15 \\
? 6202 & --- & --- & 1.67E-15 & 2.16E-15\\
FeII(74) 6248 & 1.64E-12 & ---& ---& ---\\
$[OI]$(1) 6300 & 1.15E-11  & 1.61E-13 & 4.35E-14 & 4.50E-14 \\
$[OI]$(1) 6364 & 5.12E-12 & 6.62E-14 & 2.43E-14 & 2.33E-14 \\
NII(8) 6482+FeII(74) 6456 & 6.55E-12 & 7.19E-14 & 1.52E-14 & 1.34E-14 \\
H$\alpha$ 6563 + $[NII]$(1) 6584 & 3.19E-10 & 5.26E-12 & 1.01E-12 & 1.43E-12 \\
HeI(46) 6678 & 1.02E-12 & 3.32E-14 & 6.71E-15 & 8.84E-15 \\
OI(2) 6726 & 2.11E-12 & 3.47E-14 & --- & ---\\
OI(21) 7002 & 4.50E-13 & 8.39E-15 & --- & --- \\
HeI(10) 7065 & 5.93E-13 &	7.85E-14 & 1.27E-14 & 1.65E-14  \\
CII(20) 7115/20 & 3.28E-12 & 6.20E-14 & ---& ---\\
CII(3) 7231/36  & 7.17E-12  & 8.80E-14 & 1.53E-14 & 2.02E-14 \\
$[OII]$(2) 7325 & 2.46E-12 & 1.96E-13 & 6.30E-14 & 1.09E-13 \\
NI(3) 7424/42/69 & 5.93E-13 & 3.50E-14 & --- & --- \\
OI(55) 7477/9 & 2.19E-12 & --- & --- & --- \\
CVI 7708/26 + OIV 7713 & --- & --- & 9.02E-15& 1.13E-14 \\
OI(1) 7773 & 2.19E-11 & 1.03E-13 & 8.16E-15 & 9.15E-15 \\
OI(35) 7947/48 & --- & 1.27E-14 & --- & --- \\
OI(19) 7982/87/95 & --- & 4.94E-15 & --- & --- \\
MgII(7) 8216 + NI(2) 8216 &  5.12E-12 &  1.03E-13 & 4.21E-15 & 5.49E-15  \\
MgII(7) 8238 + NI(2) 8223 & 7.58E-12 & --- & 4.77E-15 & 6.67E-15 \\
OI(4) 8446 & 1.15E-10 & 2.09E-12 & 5.28E-14 & 5.491E-14 \\
CaII(2) 8542 & 5.73E-12 & --- & --- & --- \\
NI(1) 8686/3/703 + CaII(2) 8662 & 5.25E-11 & 1.122E-13 & 7.050E-15 & 1.396E-14 \\
NI(8) 8629   & 	--- & 1.022E-13 &--- &--- \\
NI(1) 8724  & 	---	&	5.558E-14 & ---& ---\\
 HP 8751 & --- & --- & 3.754E-15 & 5.122E-15 \\
HP 8863 & --- & 4.604E-14 & 5.240E-15 & 8.226E-15 \\
? 8930 & --- & --- & 1.159E-15 & 2.686E-15  \\
HP 9015 & --- & 5.89E-14 & 6.184E-15 & 1.117E-14 \\
HP 9229 & --- & 1.51E-13 & 2.137E-14 & 3.139E-14 \\
NI(7) 9393 + CI(9) 9406	& --- &	9.93E-14 & --- & --- \\
\hline
\end{tabular}
\end{center}
\end{table*}

   \begin{figure*}
   \centering
   \rotatebox{-90}{\includegraphics[width=14cm]{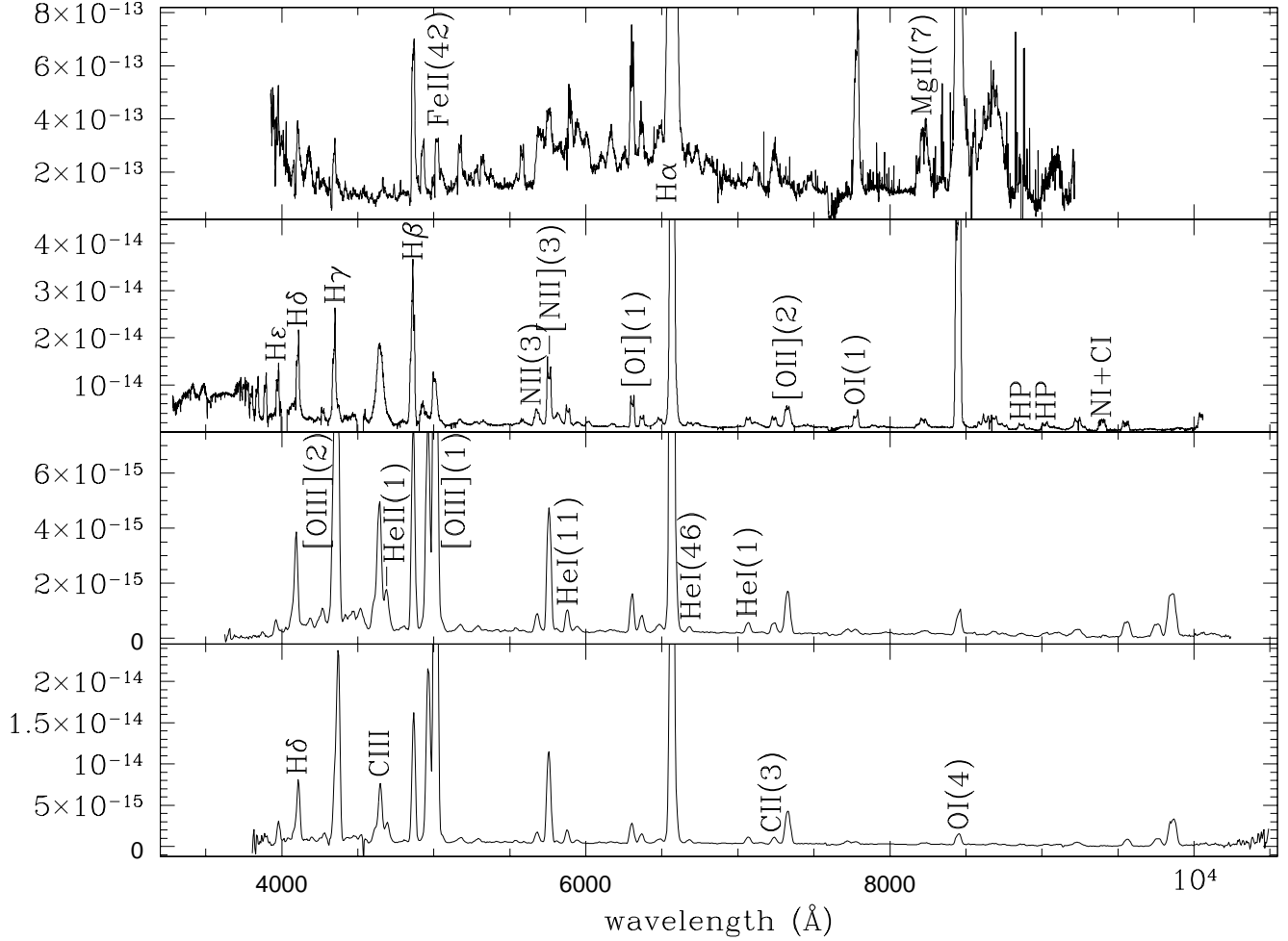}}
      \caption{Nova SMC 2001. From top to bottom: November 2001, December 2001, February 2002, and March 2002 spectra. 
Flux units are erg sec$^{-1}$cm$^{-2}$\AA$^{-1}$.}
         \label{f2}
   \end{figure*}

\begin{table*}
\begin{center}
\scriptsize
\caption{Nova LMC 2002 emission lines fluxes. Flux measurements are affected by errors of he order of $\sim 30\%$. }
\begin{tabular}{p{5cm}p{3cm}p{3cm}p{3cm}}
 & & & \\
\hline
Line & 09-03-2002 & 14-03-2002 & 11-08-2002 \\
\hline
& & &\\
H11(2) 3771 & 1.03E-12 & 4.40E-13 & ---\\
H10(2) 3798 & 4.80E-13 & 2.58E-13 & ---\\
H$\eta$ 3835 + MgI(3) 3835 & 8.34E-13 & 4.91E-13 & ---\\
H$\zeta$ 3889 & 6.24E-13 & 3.19E-13 & ---\\
CaII(1) 3934  & 2.34E-12 & 1.08E-12 & ---\\
H$\epsilon$ 3970 + CaII(1) 3969 & 2.32E-12 & 7.95E-13 & 9.40E-16 \\
$[SII]$(1) 4069 & --- & --- & 1.59E-15 \\
 ? 4091 & --- & --- & 5.65E-15 \\
H$\delta$ 4102 & 2.60E-12 & 8.24E-13  & 5.80E-15 \\
$[FeII]$(21) 4177 & --- & --- & 7.33E-16 \\
FeII(27) 4176+(28) 4179  & 9.11E-13 & 4.53E-13 & ---\\
OII(36) 4190/86 & --- & --- & 6.33E-16 \\
FeII(27) 4233 & 7.01E-13 & 4.56E-13 & ---\\
CII(6) 4267 & --- & --- & 8.92E-16 \\
FeII(27) 4273 & 4.51E-13 & 5.62E-13 &--- \\
FeII(27) 4303 + (28) 4297 & 1.10E-12 & 6.70E-13 & ---\\
H$\gamma$ 4341 & 4.30E-12 & 2.19E-12 & 4.72E-15 \\
$[OIII](2)$ 4363 & --- & --- & 1.94E-14 \\
FeII(27) 4417 & 5.47E-13 & 3.33E-13 &--- \\
OII(5) 4452 & 3.38E-13 & 1.51E-13 &--- \\
FeII(37) 4489/91 & 5.54E-13 & 4.55E-13 & ---\\
FeII(38) 4523 + (37) 4520/34 & 6.68E-13 & 6.72E-13 & ---\\
FeII(37) 4556 + (38) 4550 & 1.22E-12 & 6.93E-13 & ---\\
FeII(38) 4596 + OII(5) 4591 & 8.57E-13 & 6.850E-13 &--- \\
NII(5) 4607/14/21 & --- & --- & 1.60E-15 \\
NIII(2) 4640 & 6.78E-13 & 4.82E-13  & 7.21E-15 \\
HeII(1) 4686 & --- & --- & 2.36E-15 \\
H$\beta$ 4861 & 9.52E-12 & 6.41E-12 & 1.32E-14 \\
FeII(42) 4924   & 1.79E-12 & 1.43E-12 & --- \\
$[OIII]$(1) 4959 & --- & --- & 2.79E-14 \\
$[OIII]$(1) 5007 & --- & --- & 8.31E-14 \\
FeII(42) 5018   & 1.64E-12 & 1.50E-12 & ---\\
HeI(47) 5048  & 5.07E-13 & 5.25E-13 & ---\\
FeII(42) 5169   & 1.78E-12 & 1.29E-12 & ---\\
NII(66) 5176/80 & --- & --- & 3.24E-16 \\
FeII(49) 5198  & 1.06E-12 & 5.485E-13 & ---\\
FeII(49) 5235 +(48)  & 8.50E-13 & 4.58E-13 & ---\\
FeII(49) 5276 +(48)  & 1.16E-12 & 7.05E-13 & ---\\
FeII(49) 5317 +(48)  & 1.47E-12 & 9.42E-13 & ---\\
FeII(55) 5535 & 4.87E-13 & 3.60E-13 & ---\\
$[OI](3)$ 5577 & 5.78E-13 & 5.14E-13 & 6.75E-16 \\
NII(3) 5667/80 & 4.62E-13 & 5.050E-13 &  1.47E-15\\
$[NII](3)$ 5755 & 3.55E-13 & 3.66E-13 & 1.50E-14 \\
CIV(1) 5805 & 3.13E-13 & 2.29E-13 & 7.28E-16 \\
HeI(11) 5876 & --- & --- & 1.71E-15 \\
NaI(D) 5896 & 7.88E-13 & 5.85E-13 & --- \\
NII(28) 5940/42 & --- & --- & 6.47E-16 \\
FeII(74) 6147 + NII(36/60) 6168  & 8.23E-13 & 5.32E-13 &--- \\
FeII(74) 6248 & 4.95E-13 & 3.36E-13 &--- \\
\end{tabular}
\end{center}
\end{table*}

\begin{table*}
\begin{center}
\setcounter{table}{5}
\scriptsize
\caption{continued}
\begin{tabular}{p{5cm}p{3cm}p{3cm}p{3cm}p{3cm}}
$[OI]$(1) 6300 & 6.29E-13 & 7.07E-13 & 6.95E-15 \\
$[OI]$(1) 6364  & 4.03E-13 & 3.29E-13 & 2.80E-15 \\
FeII(74) 6417 & 2.53E-13 & --- & --- \\
FeII(74) 6456 & 6.92E-13 & 3.546E-13 & --- \\
NII(8) 6482 & --- & --- & 1.37E-15 \\
H$\alpha$ + [NII](1) 6584 & 2.88E-11 & 2.83E-11 & 8.33E-14 \\
HeI(46) 6678 & --- & --- & 4.69E-16 \\
HeI(10) 7065 & --- & --- & 1.32E-15 \\
CII(20) 7115/9 & 1.96E-12 & 7.85E-13 & ---\\
CII(3) 7234 & 7.75E-13 & 6.32E-13 & 2.06E-15 \\
$[OII]$(2) 7319/31 & 3.35E-13 & 7.93E-14  & 1.32E-14 \\
OI(55) 7477/79/81 + NI(3) 7424/42/68  & 1.76E-12 & 4.72E-13 & --- \\
CIV 7726 & --- & --- & 1.67E-15 \\
OI(1) 7773  & 5.66E-12 & 3.22E-12 & 5.71E-16 \\
MgII(8) 7877/86 & 1.08E-12 & --- & ---\\
OI(34) 8227 & 7.20E-13 & 4.73E-12 & ---\\
CaII(13) 8250 & 1.33E-12 & 6.68E-13 & --- \\
HP(11) 8346 + CI(10) 8335 & 1.10E-12 & 4.55E-13 & ---\\
OI(4) 8446/7 & 4.94E-12 & 4.77E-12 & 2.35E-15 \\
CaII(2) 8498 & 3.08E-12 & 1.977E-12 & ---\\
CaII(2) 8542 & 2.85E-12 & 2.05E-12 & ---\\
CaII(2) 8662 &  2.18E-12 & 1.56E-12 & ---\\
$[CI]$ 8727 & 2.71E-12 & 1.49E-12 & ---\\
HP(9) 8863 & 1.16E-12 & 5.71E-13 & ---\\
NI(15) 9061 & 2.95E-12 & 1.10E-12 & \\
CI(3) 9087 & 5.40E-12 & 2.86E-12 & --- \\
MgII(1) 9226 & 1.73E-12 & 1.12E-12 & --- \\
OI(8) 9264 & 2.47E-12 & 1.13E-12 & --- \\
CI(9) 9406 & 2.91E-12 & 2.03E-12 & --- \\
HP(8) 9546 & 7.92E-13 & 5.39E-13 & --- \\
CI(2) 9627 & 1.39E-12 & 7.87E-13 & --- \\
CI(2) 9658 & 1.81E-12 & 8.59E-13 & --- \\
HP(8) 10049 & 1.89E-12 & 1.33E-12 & --- \\
NI(18)  10117 & 2.14E-12 & 1.29E-12 & --- \\
\hline
\end{tabular}
\end{center}
\end{table*}

   \begin{figure*}
   \centering
\rotatebox{-90}{\includegraphics[width=10cm]{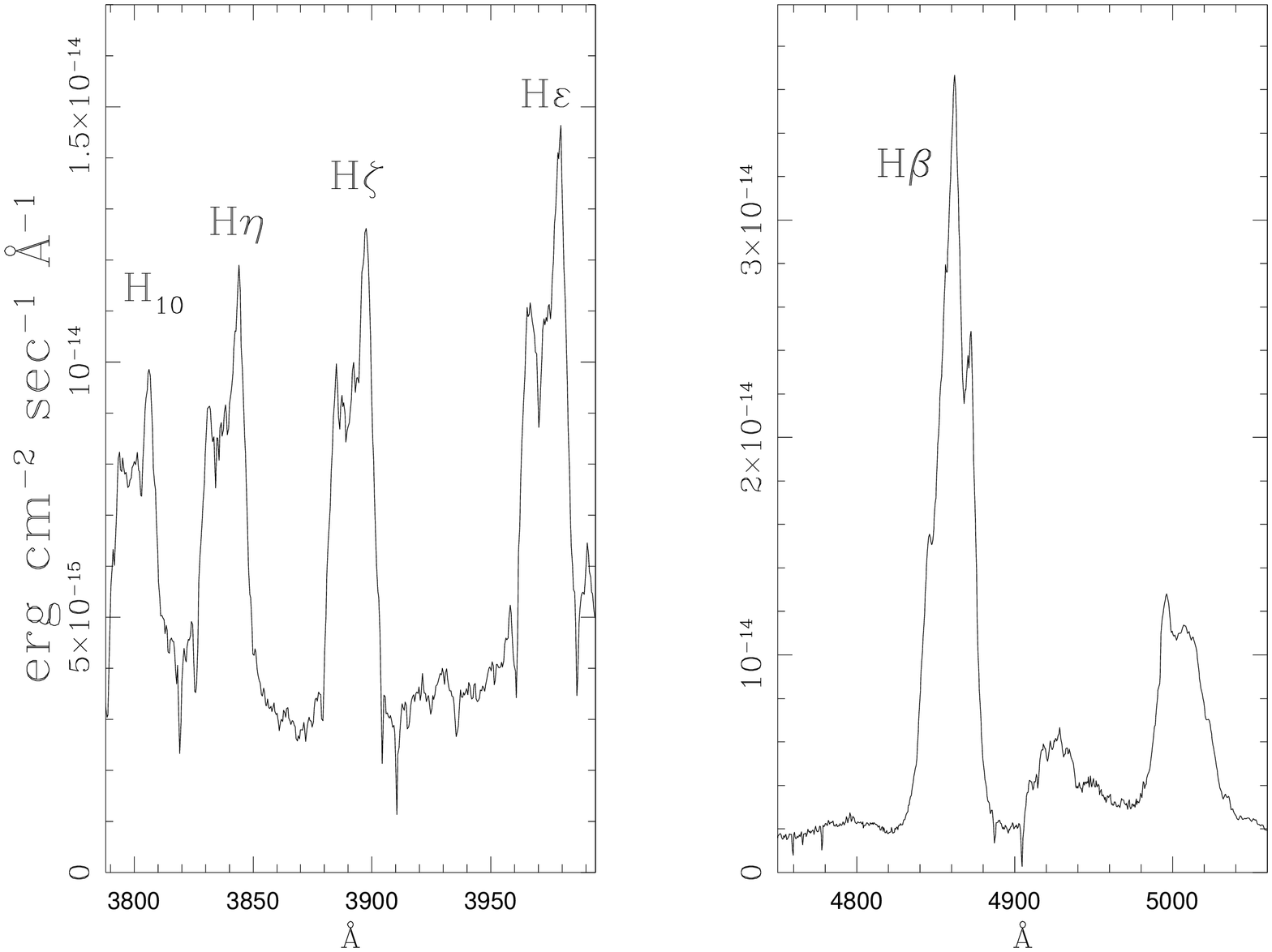}}
\caption{Zoomed-in view of Nova SMC 2001 December spectrum. The two panels show the different profiles of the Balmer lines 
(see also Sec.~3.1). }
   \end{figure*}

\section{Spectral analysis}

\subsection{The observed evolution}

We identified the emission lines and measured the corresponding fluxes in
order to classify the novae according to the Cerro Tololo system
(Williams et al. 1991, 1994), and to
provide input parameters for the
computation of the elemental abundances (e.g., Shore et al. 2003). The
observed spectroscopic evolution of each nova is described below and
summarized in Table~3.

For both novae we estimated the V band magnitude at each epoch (See Table~4)
by convolving the observed spectra with the V-Johnson filter.
The synthetic magnitude of Nova SMC 2001 obtained on March 4 (2002)
may indicate that the nova was undergoing a re-brightening of at least one
magnitude, thus suggesting the formation of dust within the ejecta
similar to FH~Ser, NQ~Vul, and Nova Aql 1982 (Hack et al. 1993).
However, since the sky conditions during the observation were not reported,
we cannot rule out that the sky was not photometric,
and that the re-brightening is an artifact of the calibration.
The line flux measurements for Nova SMC 2001 and Nova LMC 2002 are given in
Table~5 and 6, respectively. \\

\noindent{{\it Nova SMC 2001}}

The spectroscopic follow-up of Nova SMC 2001 covers days 11, 46, 119,
and 134 since the observed maximum (Fig.~2). 

The spectrum taken in
November 2001 (11 days past maximum) is dominated by the Balmer lines
and OI $\lambda$8446. We also observed permitted transitions from both
low (FeII, NaI, and MgII) and moderate excitation lines (HeI, CII, and
NII).  Forbidden transitions from [OI] and [NII] have been detected,
too.  The Balmer lines H$\beta$, H$\gamma$ and H$\delta$ show double
P-Cyg absorptions whose minima indicate average expansion velocities
of --1300 and --1000 km/sec.  The FeII (42) emission lines show just a
single P-Cyg absorption, indicating an average expansion velocity of
$\sim-$900 km/sec.

In December 2001 (46 days past maximum)
Nova SMC 2001 exhibited the $[OIII]$ lines
$\lambda\lambda$4959,5007 which
characterizes the ``nebular phase'' of novae (Payne-Gaposchkin 1957).
The line profiles are clearly all saddle shaped with the red component
(R) stronger than the blue one (B), with the exception of the H$\beta$ line
for which  B$>$R (see Fig.3).
The separation between the red and the blue component
is $\sim$900$\div$1000 km/sec, whereas, the FWHM of the emission lines is
$\sim$1400$\div$1900 km/sec. Saddle shaped profiles are suggestive of an
expanding equatorial shell or ejection of polar caps (Payne-Gaposchkin 1957).

The spectra taken on February and March 2002 (119 and 164 days past maximum, respectively) are quite similar and are dominated by strong auroral and nebular emission lines such as
 [OIII]$\lambda$5007 (as strong as the H$\alpha$ emission), and
[OIII]$\lambda$4363.
All the permitted lines have considerably weakened,
while the ``4640'' blend is now
clearly resolved into NIII $\lambda$4640, HeII $\lambda$4686, and NV
$\lambda$4609. On the basis of
our spectroscopic observations we classify the spectra
obtained in November and December 2001 as
$P_{na/n}^o$ and $P_n^o$, respectively,
and those taken in  February and March 2002
 as $A_o$, according to the Cerro Tololo scheme. \\

\noindent{{\it Nova LMC 2002}}

We obtained spectra of nova LMC 2002 6, 11 and 161 days after
 discovery (see Fig.~4 and Table~6).  From the light curve in Fig.~1 we
derive $t_2=12$ days, which makes Nova LMC 2002 a borderline object
between the {\it fast} and {\it slow} nova classes as defined by Della
Valle \& Livio (1998).

The early spectra (6 and 11 days past maximum) are dominated by
 Balmer and low ionization emission lines such as FeII, OI, CaII, MgII and
NaI D. Weak emission lines from  CI, CII and NI are
 detected red-ward of $\sim$7000\AA. We also observe the [OI]
$\lambda$5577 and
  $\lambda\lambda$6300,6364 emission lines.
These spectra are quite
 similar to those of Nova LMC 1988 No.1 three days after maximum, and
 Nova Scuti 1989 fifteen days after maximum. We classify our
 spectra as P$_{fe}$ according to the Cerro Tololo system.

The spectrum  taken in August 2002 (161 days past maximum)
is dominated by strong auroral and
nebular lines from [OIII] and [NII]. All low ionization emission lines
 have completely disappeared, and the only permitted transitions are due to
NII, HeII, CII, NIII, CIV, OI and OII.
At this stage the spectrum corresponds to the $A_o$
phase in the Cerro Tololo system.
This fact, coupled with an expansion velocity of
2150 km/sec, as measured from the $\langle HWZI \rangle$ of the Balmer
lines near maximum, indicates that Nova LMC 2002 belongs to the {\it FeII}
spectroscopic class (indeed HWZIs smaller than 2500 km/sec are typical of
{\it FeII} type novae, see Williams 1992).

   \begin{figure*}
   \centering
   \rotatebox{-90}{\includegraphics[width=14cm,]{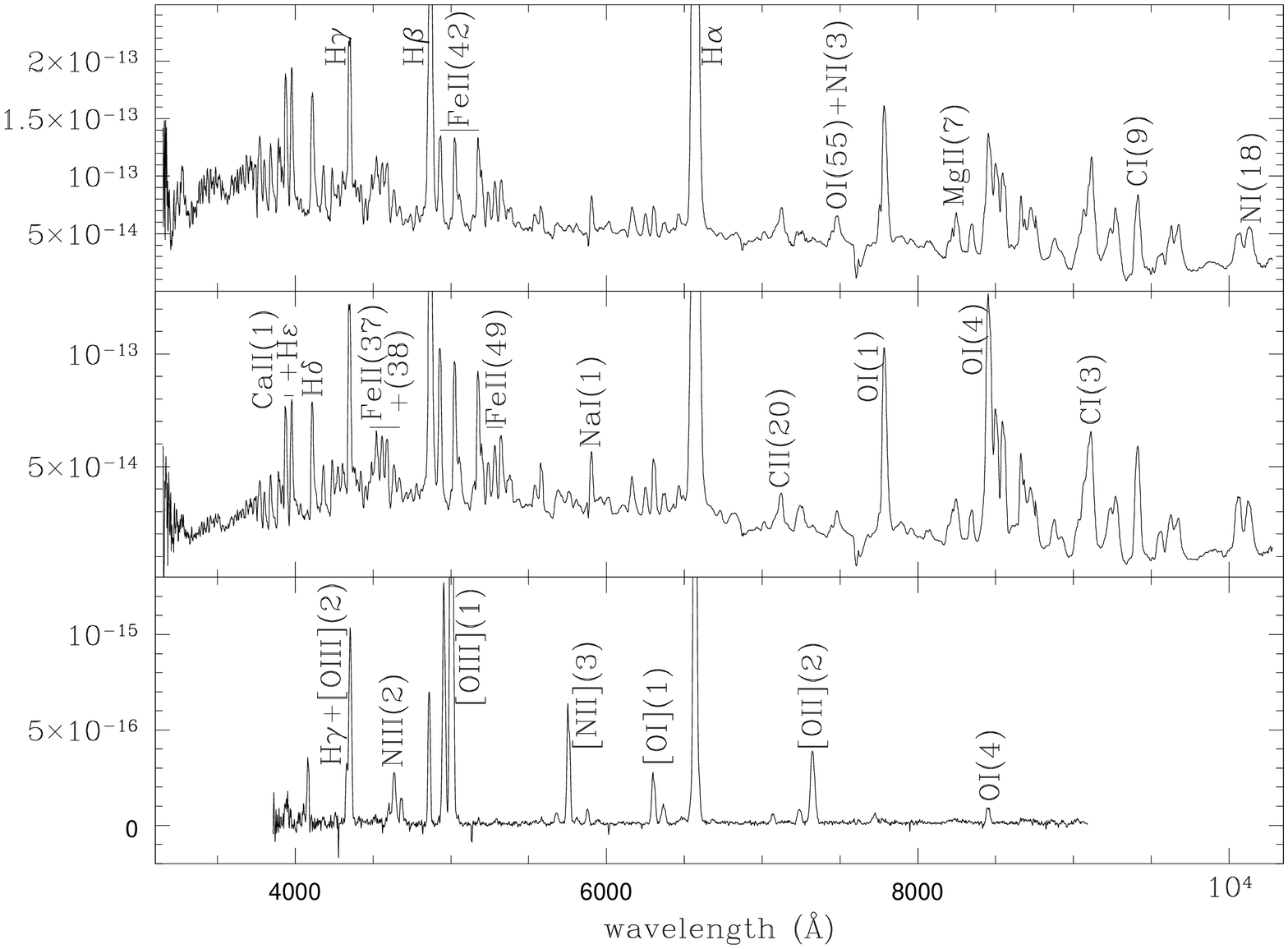}}
      \caption{Nova LMC 2002 spectra. From top to bottom: March 9, March 14,
and August 8 spectra. Flux units are erg sec$^{-1}$ cm$^{-2}$ \AA$^{-1}$. }
         \label{f4}
   \end{figure*}

\begin{table*}
\begin{center}
\scriptsize
\caption{The OI line flux ratio, the corresponding opacity
$\tau_{\lambda6300}$, $T_e$, and mass derived for each nova at each
epoch according to Williams (1994). See text for more details. }
\begin{tabular}{cccccc}
 & & & & & \\
\hline
object & obs date & $r=\frac{F_{\lambda6300}}{F_{\lambda6364}}$ & $\tau_{\lambda6300}$ & $T_e(K)$ & $M_{OI}(M_\odot)$ \\
\hline
 & & & & & \\
Nova SMC 2001 & 01/11/01 & 2.24 & 0.98 & 4900 & 9.83e-4 \\
	& 06/12/01 & 2.43 & 0.68 & 4600 & 1.89E-5 \\
 	& 17/02/02 & 1.79 & 1.96 & 3500 & 4.30E-5 \\
	& 04/03/02 & 1.93 & 1.60 & 3600 & 3.28E-5 \\
Nova LMC 2002 & 09/03/02 & 1.56 & 2.75 & 5200 & 7.96E-5 \\
	& 14/03/02 & 2.15 & 1.14 & 5600 & 3.82E-5 \\
	& 11/08/02 & 2.48 & 0.61 & 4000 & 1.46E-6 \\
\hline
\end{tabular}
\end{center}
\end{table*}

\subsection{Reddening, flux measurements and the neutral oxygen mass}

\begin{table*}
\begin{center}
\scriptsize
\caption{The observed $[OIII]$ line ratio and the computed filling
factor, $\varepsilon$, and hydrogen mass. The spectra used are those
taken in March 2002 and August 2002 for Nova SMC 2001 and Nova LMC
2002, respectively. See section 3.3 for more details.}
\begin{tabular}{ccccccc}
 & & & & & & \\
\hline
object & (4959+5007)/4363 & $N_e$ & $T_e$ & $f$ & $\varepsilon$ & $M_H$ ($M_\odot$)\\
\hline
 & & & & & & \\
Nova SMC 2001 & 3.65 & 10$^7$ & 12000 & 0.41 & 0.16 & 6.07E-5 \\
              &      & 10$^8$ & 8500  & 0.34 & 0.002& 5.02E-6 \\
Nova LMC 2002 & 5.69 & 10$^7$ & 10000 & 0.45 & 0.01 & 2.38E-6 \\
              &      & 10$^8$ & 7600  & 0.39 & 1.0E-4 & 2.0E-7 \\
 & & & & & & \\
\hline
\end{tabular}
\end{center}
\end{table*}

The emission line fluxes of dereddened spectra at each epoch are reported
in Table 5 and 6 for Nova SMC 2001 and Nova LMC 2002, respectively.  The
correction for extinction was applied assuming
R=3.1, and E(B--V)=0.09 ($\pm$0.02) and 0.13 ($\pm$0.05) for Nova SMC
2001 and Nova LMC 2002, respectively (see also Sec.~3.4). The
values for E(B--V) were computed by averaging the extinction estimates in
the literature (Dutra et al. 2001; Oestreicher \& Schmidt-Kaler 1996;
Oestreicher et al. 1995; Massey et al. 1995; Schwering \& Israel 1991;
Bessel 1991; and Capaccioli et al. 1990). Extinction values derived from
extinction maps of the Magellanic Clouds were weighted twice
in our computation.

The calculated correction, however, does not take into account the internal
extinction of each nova, which can be higher due to dust formation. This can be checked by using the intensity ratio of
optically thin lines which have the same upper level.
We could use the H(8)/HP(8) line ratio measured in the December spectrum of Nova SMC 2001. We derived an extinction correction of E(B-V)=0.26$\pm$0.20 (cfr. Vanlandingham et al. 1999), which is indeed larger than the above value of 0.09$\pm$0.02, although still consistent within the errors.

Temperature, $T_e$, and mass of the neutral
oxygen in the ejecta have been computed following Williams (1994).
In case of relatively high densities (as in novae ejecta) the [OI] lines
$\lambda$5577 and $\lambda$6300 are easily detectable and can be used as a
diagnostic of the electron temperature. After substituting in the equation for the line flux ratio  the
 Einstein coefficients, the excitation potential
 and the statistical weight of the energy levels (see Williams 1994 and references therein) we get:
\begin{equation}
T_e=\frac{11200}{\log \left(\frac{43\tau}{(1-e^{-\tau})}\times\frac{F_{\lambda6300}}{F_{\lambda5577}}\right)}
\end{equation}
where $\tau$ is the optical depth of the line $\lambda$6300 as derived by solving the equation:
\begin{equation}
\frac{F_{\lambda6300}}{F_{\lambda6364}}=\frac{1-e^{-\tau}}{1-e^{-\tau/3}}
\end{equation}
and $F_{\lambda(i)}$ is the observed (unreddened) line flux at the specified wavelength.
The OI mass can be derived knowing that the line flux is proportional to the mass of the emitting region (the optical depth is not large) scaled by the square of the distance, $d$, i.e (again, after substitution of the tabulated constants):
\begin{equation}
\begin{array}{rcl}
M_{OI}&=&152\times d^2_{kpc} \exp\left(\frac{22850}{T_e}\right) \\[3mm]
      & &\times10^{1.05 E(B-V)}\frac{\tau}{1-e^{-\tau}}F_{\lambda6300} M_\odot\\
\end{array}
\end{equation}

The measured line
flux ratios and the derived gas temperature and mass are reported in
Table~7.  We found that the gas temperature, $T_e$, is within the range
3500-5000 K for both novae (cfr. Williams 1994, his Table~2).
The OI mass was computed assuming the distances of the LMC and SMC to
be 51.4 kpc (Panagia 1998) and 56.4 kpc, respectively, the latter value
resulting from the average of recent distance measurements
(Harries et al. 2003; Dolphin et al. 2001; Bono et al. 2001; and Cioni
et al. 2000).  We derived OI masses of the order of a few $10^{-5}
M_\odot$ (see our Table~7 and Table~3 of Williams 1994).

\subsection{The HII mass and the filling factor}

We estimate the mass of the ionized hydrogen and its filling factor,
following the approximation described in Mustel \& Boyarchuk
(1970)\footnote{We used the H$\beta$ flux and not the H$\alpha$ flux as in
Mustel \& Boyarchuk (1970), due to the heavy blend of the H$\alpha$
and $[NII]$ lines.}. The volume of the ejecta can be approximated
by a spherical shell of radius $r=v_{exp}\Delta t$ (where $v_{exp}$ is
the expansion velocity measured from the spectra and $\Delta t$ is the
time since maximum), and thickness $fr$, with $f$ given by the ratio
of the gas thermal velocity over the expansion velocity,
i.e. $v_{therm}/v_{exp}$. The volume is furthermore corrected by a
factor $\varepsilon \leq1$, the filling factor, which parametrizes the
clumpiness of the ejecta. Thus:
\begin{equation}
V=4\pi (v_{exp}\Delta t)^3 f\varepsilon
\end{equation}

The observed line flux is the emitted flux reduced by the inverse
square effect of the geometrical dilution, thus:
\begin{equation}
I_\lambda=\left[\left(\frac{4\pi j_\lambda}{N_e N_p}\right) N_e N_p  V\right]\frac{1}{4\pi d^2}
\end{equation}
where $I_\lambda$ is the measured line flux, $j_\lambda$ is the gas
emissivity, the factor $4\pi j_\lambda/(N_pN_e)$ is tabulated in
Osterbrock (1989),  $N_p$ ($N_e$) is the proton (electron)
density, and $d$ is the distance.  Knowing the volume,
the electron density of the ejecta, and assuming $N_p\sim N_e$, we can
compute the HII mass; furthermore, combining the equations (4) and (5) we
can derive the filling factor.
Our spectra do not show any of the typical lines used as a diagnostic
for the electron density (see Osterbrock 1989), thus, we have proceeded
according to Filippenko \& Halpern (1984). Using the observed [OIII]
line ratio (4959+5007)/4363, we derived the electron density, $N_e$,
corresponding to the typical nebular temperature of $\sim10^4$K. We
found that a density of $N_e\sim10^7$ cm$^{-3}$ is required in both
the Nova SMC March 2002 spectrum and Nova LMC August 2002 spectrum. In order
to obtain temperature values which are close to those reported in Table~7
(namely 3500$\div$5000 K), we would need input densities higher than $10^8$ cm$^{-3}$. However, such high densities are hardly compatible with nebular
regimes. In addition, the observed [OI] $\lambda$6300/$\lambda$6364
line ratio (Table~7, col.~3) is
definitely smaller than the theoretical value of $\sim$3, expected at nebular
regimes (Osterbrock 1989).  This suggests that the [OI] and the Balmer+[OIII]
emissions arise from different ejected layers.
Williams
(1994) has explained this behavior (exhibited by most novae) by
assuming that the optically thick lines of [OI] originate in cool, high
density blobs ($N(H)>10^{14}$ cm$^{-3}$) of neutral material embedded within
the ejected shell. The filling
factor measurements $\varepsilon \sim 10^{-4}\div 10^{-1}$ (Table~8,
col. 6) lend support to this view, thus suggesting that both Nova SMC
2001 and Nova LMC 2002 have clumpy ejecta rather than shells
homogeneously filled in.

\subsection{The absolute magnitudes}

The study of the distribution of the absolute magnitude of
novae at maximum in the MC has a specific interest.
Indeed, several authors (e.g., van den Bergh and
Pritchet 1986; Capaccioli et al. 1989) have already pointed out
 the existence of a
subclass of superluminous classical novae which are systematically
brighter (by about 1 mag) than predicted by the general  MMRD (Maximum Magnitude vs. Rate of Decline) relation (Della Valle and Livio 1995). The last
example of such an object was the exceptionally bright Nova LMC 1991
(Della Valle 1991).
 The frequency of occurrence of these events is very uncertain
and its estimate is of interest: on one hand they are
potential `caveats' for using the Novae as distance indicators in very
distant galaxies (e.g., Pritchet and van den Bergh 1987; Della Valle and
Gilmozzi 2002). On the other hand they can play an important
role in understanding the nature of the nova phenomenon
(see Schwarz et al. 2001). In the case of Nova LMC 2002, both t$_2=12$ days
and the apparent magnitude at maximum, $m_V=10.4$ mag, are consistent
with the values of a `normal' nova. Furthermore, we have in hand enough
parameters to provide an independent check. From $t_2$=12 days, one
infers an absolute magnitude at maximum of $M_V=-8.6$ through Della
Valle \& Livio's (1995) MMRD relation.  We then used the observed
apparent maximum magnitude, $m_V=10.4$ mag, and the LMC distance of
51.4 kpc (Panagia 1998), to solve the equation of the distance modulus
for $A_V$. We have found $A_V=0.42$ and E(B--V)=0.14, in excellent
agreement with the average $\langle E(B-V) \rangle$=0.13, derived
independently and adopted through this paper.

Nova SMC 2001 was discovered during the early decline, thus we cannot
repeat the above cross-check, but rather work out the nova
absolute magnitude in a different way. Taking advantage of the
similarities between nova SMC 2001 and nova LMC 1988 No.1 (see Sec.~1),
we can assume $t_2<20$ days for nova SMC 2001 (see also Table~9). This
 implies that the nova was about 1 mag below maximum at the time of
Liller's discovery. The absolute maximum magnitude of Nova SMC 2001 is
thus constrained within the range $-8.3 < M_V < -7.7$ mag.

\subsection{The nova populations}

\begin{table*}
\begin{center}
\scriptsize
\begin{tabular}{lp{4cm}ccc}
 & & & & \\
\hline
nova & spectral description.& spectral class & t2 & ref \\
\hline
 & & & & \\
LMC 2002 & HWZI$\sim$2150, FeII and low ionization elements & FeII & 12 & this paper \\
LMC 1995$^\dagger$     & P-Cyg profiles (900 and 1500 km/sec), Fe and Na emission lines early dust formation & FeII & 7 & 1, 2\\
LMC 1992     & very narrow emission lines, P-Cyg profiles, Na and Fe & FeII & 6 & 3\\
LMC 1991     & FWHM 2300 km/sec & FeII & 6 & 4, 5\\
LMC 1990 \#2 & strong HeI and HeII lines, FWHM$\sim$5500 km/s, mimic 1968 outburst and RN U~Sco and  V394~CrA & He/N & 3 & 6, 7\\
LMC 1990 \#1   & very broad emission lines (FWHM$\sim$5600 km/s), UV maximum spectrum very similar to V693~CrA and  U~Sco. Early development of HeI and HeII emission lines as well as [Ne]. & He/N & 4.5 & 8, 9\\
LMC 1988 \#2 & broad Balmer and FeII emission lines, P-Cyg profiles, FWHM$\sim$3500 km/sec & FeIIb & 5 &  10, 11\\
LMC 1988 \#1 & FeII, NaI, OI and CaII with P-Cyg profiles &  FeII & 20  & 10, 12 \\
LMC 1981$^{\dagger\dagger}$     & very fast and similar to V1500 Cyg & FeIIb  & 3  &13 \\
LMC 1978 \#1     & narrow Balmer lines, {\it normal} past maximum spectrum, fast decline & FeII & 8 & 14, 15\\
\hline
\end{tabular}

\caption{
LMC novae in the literature. References: IAUC 6143 (1); AAVSO
interactive light curve (2); IAUC 5651, 5653, 5656, 5657, 5659, 5661,
5669, 5683 (3); IAUC 5260 (4); Schwartz et al. 2001 (5); IAUC 4964,
4975 (6); Sekiguchi et al. 1990 (7); IAUC 4946, 4949, 4956, 4960,
4961, 4964 (8); Vanlandingham et al. 1999 (9); Williams et al. 1991
(10); IAUC 4663, 4664, 4666, 4670, 4673 (11); IAUC 4568, 4569, 4574,
4577, 4580, 4585, 4588, 4589, 4601, 4610 (12); IAUC 3648, 3641 (13),
IAUC 3206 (14), Capaccioli et al. 1990 (15).
$^\dagger$ Nova LMC 1995's $t_2$ is extrapolated assuming linear rate
of decline from IAUC data points covering the first 4 days. A similar
value may possibly be inferred from AAVSO unvalidated data points on
line.
$^{\dagger\dagger}$ The $t_2$ for nova LMC 1981 is unknown:
Duerbeck (1981, IAUC 3641) reports that the nova is very fast and that
the spectrum taken few days after discovery strongly resemble the
spectra of CP~Pup and V1500~Cyg +4.5 mag from maximum. In such a
hypothesis we assume $t_2\sim$3-5 as given in the table above. On the
other side Maza (1981, IAUC 3641) confirming the discovery of the nova
after 6 days from the first announcement reports a decline of +0.7
mag. As rate of decline is roughly linear within 2 or 3 mag from
maximum, and assuming the decline rate given by Maza the nova should
have a $t_2$ of $\sim$18 days.  }
\end{center}
\end{table*}

 We collected from the literature all the photometric and  spectroscopic
data for LMC novae in order to derive the $t_2$ from the light curves and the
spectroscopic classification according to the Cerro Tololo scheme (see Table~9). We used the data reported in Table~9 together with those in Table~1 and 2 of Della Valle and Livio (1998), to plot the
frequency distribution (Fig.~5) of the rates
of decline for the novae in the LMC and the Milky Way (Fig.~5, lower and upper panel, respectively).

The distribution of the nova population in the Milky Way is  bimodal, and
consists of {\it fast} declining objects ($\log(t2)<1.1$) originating mostly in the disk and
{\it slow} declining novae forming also in the bulge/thick disk (Della Valle et al. 1992; Della Valle and Livio 1998).
The observed LMC novae distribution (bottom panel of Fig.~5) matches the disk
population within the Milky Way (top panel of Fig.~5). This is an expected result since the LMC is a bulge-less galaxy. Moreover,
  Subramaniam \& Anupama
(2002), studying the parent population of the novae in the LMC, found
that LMC classical novae are surrounded by an intermediate age population within
the range 3.2$\div$1.0 Gyr.
 Subramaniam \& Anupama (2002) studied also the parent population around Nova SMC 1994 and found that Nova SMC 1994, which has the same spectroscopic evolution as Nova SMC 2001 (see Sec.~4), is surrounded by an old population (1$\div$10 Gyr) similar to the two slow novae
observed in the LMC.

\section{Summary and Conclusions}

In this paper we have presented  the early-decline/middle-stages
spectra of two recent novae discovered in the Magellanic clouds:
Nova SMC 2001 and Nova LMC 2002.

Nova SMC 2001 is the first nova in the SMC that has been
spectroscopically observed during the early decline. This object
displays a classical {\it FeII}-type spectrum, characterized by
moderately high expansion velocities (FWHM $\sim$1900 km/sec).
The observations obtained during the nebular phase indicate
that Nova SMC 2001 developed a {\it
standard} nebular spectrum (Williams 1992). The
auroral-nebular phase of Nova SMC 2001 is similar to that exhibited by
Nova SMC 1994, the only other SMC nova which has been
spectroscopically observed (though only during its nebular phase, de~Laverny et al. 1998).

Nova LMC 2002 is characterized by a rapid photometric evolution,
$t_2\sim12$ days, and a relatively high expansion velocity (HWZI=2150 km/s). Both
aspects make this nova a borderline object  between the
{\sl fast} and {\sl slow} nova classes defined by Della Valle and Livio (1998).
The spectroscopic evolution shows a typical
{\it FeII} type object and our latest observation suggests
the development of a {\it standard nebular spectrum}.

Spectra of the two novae were analyzed to provide emission line
fluxes, measurements of the OI and HII ejected masses (Table~7 and 8),
and estimates of the filling factor (Table~8). We obtained masses of
the order of {\sl a few} $\times 10^{-5}$M$_\odot$ and $\sim
10^{-6}$M$_\odot$, for the OI and the HII, respectively. After
assuming that the O mass represents $\sim$15\% of the total mass of
the shell (Williams 1994 and Gehrz et al. 1993), we can infer that
the mass of the ejecta is of the order of $2\div 3\times
10^{-4}$M$_\odot$.  We estimated the filling factor of the ejecta to
be in the range $\varepsilon\sim 10^{-4}\div 10^{-1}$. In spite of the
rough assumptions used in our computations, the derived values are
consistent with the estimate ($\varepsilon\sim 10^{-5}\div 10^{-2}$)
provided by Shara et al. (1997), after studying the ejecta of the
recurrent nova T~Pyx on {\sl HST} images.

The intensity ratio [OI] $\lambda6300$/$\lambda6364$ shows an anomalous
value of $r\sim 2$ rather than $\sim
3$ as expected on theoretical grounds. We note that such an anomalous
ratio is commonly observed in Milky Way novae (e.g. Nova V382 Vel
1999, Della Valle et al. 2002). The optically thick [OI] lines are likely
formed in high density blobs of neutral material (see
Williams 1994 for a discussion), thus suggesting a clumpy structure
of the nova ejecta. A clumpy ejecta is also suggested by the values
$\varepsilon << 1$ reported above.

Our results indicate that some of the general physical properties of
LMC and SMC novae are comparable to those of galactic novae
(e.g. spectral evolution, clumpiness of the ejecta, etc). The main
difference between the galactic and LMC nova population is in the
distribution of $t_2$ (see Fig.~5 ). The LMC nova population is formed
by bright and fast novae, whereas the faint-slow component, which is
observed in the Milky Way, is completely missing.  This behavior is
not due to an observational bias (see Della Valle 2002 for a
discussion) and it can be explained, at least in part, by theoretical
models (Starrfield et al. 1998, 2000), which predict a larger
brightness and a larger amount of accreted mass at low metallicity
regimes.

   \begin{figure}
   \centering
   \rotatebox{-90}{\includegraphics[width=7cm,]{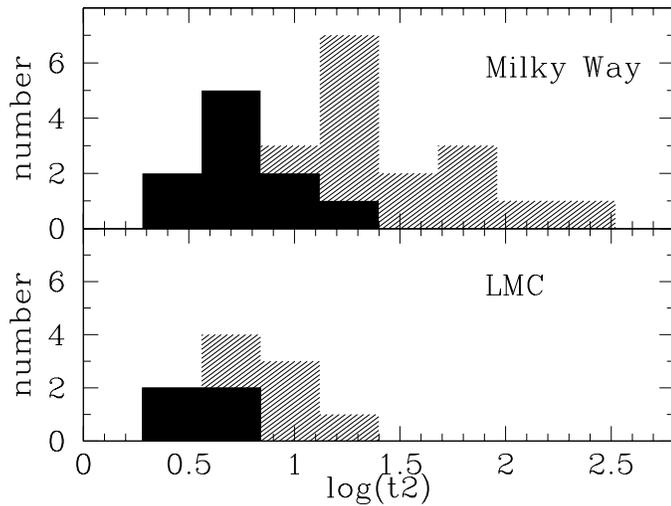}}
      \caption{Histogram of the classical nova distribution versus decline speed for the Galaxy (top panel) and the LMC (bottom panel). Dark area represents the  He/N and FeIIb classes; shaded area is for FeII novae.  See text for discussion.  }
         \label{f5}
   \end{figure}


\begin{thebibliography}{}


\bibitem[1991]{bessel} Bessel, M. S., 1991, A\&A, 242, L17
\bibitem[2001]{gb} Bono, G., Caputo, F., Marconi, M., 2001, MNRAS, 325, 1353
\bibitem[2001]{bosch} Bosch, G. L., Barba, R. H., Morell, N. I., 2001, IAUC N.7744
\bibitem[1989]{cappa1} Capaccioli, M., Della Valle, M., Rosino, L., D'Onofrio, M., 1989, AJ, 97, 1622
\bibitem[1990]{cappa} Capaccioli, M., Della Valle, M., D'Onofrio, M., Rosina, L., 1990, ApJ, 360, 63
\bibitem[2000]{cio} Cioni, M.-R.C., van der Marel, R. P., Loup, C., Habing, H. J., 2000, A\&A, 359, 601
\bibitem[1998]{lave} de Laverny, P., Beaulieu, J. P., Asplund, M., Kilkenny, D.,Renault, C., Ferlet, R., Marquette, J. B., Vidal-Madjar, A., Maurice, E., Prevot, L., and 21 coauthors, 1998, A\&A, 335, 93L
\bibitem[1991]{dv} Della Valle, M., 1991, A\&A, 252, L9
\bibitem[1992]{dv2} Della Valle, M., Bianchini, A., Livio, M., Orio, M., 1992, A\&A, 266, 232
\bibitem[1994]{ros} Della Valle, M., Rosino, L., Bianchini, A., Livio, M., 1994, A\&A, 287, 403
\bibitem[1995]{dv} Della Valle, M., Livio, M., 1995, ApJ, 452, 704
\bibitem[1989]{dl} Della Valle, M., Livio, M., 1998, ApJ, 506, 818
\bibitem[2001]{dv} Della Valle, M., Saviane, I., Williams, R. E., 2001, IAUC N. 7743
\bibitem[2000]{nconf} Della Valle, M., 2002, {\it Classical nova explosions}, AIP conference proceeding, 637, p. 443
\bibitem[2002]{dv4} Della Valle, M., Pasquini, L., Daou, D., Williams, R. E., 2002, A\&A, 390, 155
\bibitem[2002]{dvg} Della Valle, M., Gilmozzi, R., 2002, Science, 296, 1275
\bibitem[2001]{dl} Dolphin, A. E., Walker, A. R., Hodge, P. W., Mateo, M., Olszewski, E. W., Schommer, R. A., Suntzeff, N. B., 2001, ApJ, 562, 303
\bibitem[1990]{due} Duerbeck, H. W., 1990, in IAU Coll. 122, Physics of Classical Novae, eds. A. Cassatella, \& R. Viotti Springer\& Berlin, p. 34
\bibitem[2001]{d} Dutra, C. M., Bica, E., Clari$\acute{a}$, J. J., Piatti, A. E., Ahumada, A. V., 2001, A\&A, 371, 895
\bibitem[1984]{fili} Filippenko, A. V., Halpern, J. P., 1984, ApJ, 285, 474
\bibitem[1998]{gw} Gehrz, R. D., Truran, J. W., Williams, R. E., 1993, in {\it Protostar and planets III},  E.H. Levy \& J.I.Lunine
eds., Tucson: Univ. Arizona Press, p. 75
\bibitem[2002]{g} Gilmore, A. C., 2002, IAUC N.7853
\bibitem[1993]{hack} Hack, M., Salvelli, P.L., Duerbeck, H., 1993, in {\it Cataclysmic variables and related objects}, NASA, p.261
\bibitem[2003]{hr} Harries, T. J., Hilditch, R. W., Howart, I. D., 2003, MNRAS, 339, 157
\bibitem[2001]{jen} Jensen, E. L. N., Allen, P., Schwarz, G. J., Vanlandingham, K. M., 2001, IAUC N.7743
\bibitem[2002]{keg} Kilmartin, P., Gilmore, A. C., 2002, IAUC N.7847
\bibitem[2001]{l01} Liller, W., 2001, IAUC N.7738
  \bibitem[2002]{lil} Liller, W., 2002, IAUC N.7841
\bibitem[1995]{m} Massey, P., Lang, C. C., De Gioia Eastwood, K., Germany, C., 1995, ApJ, 438, 188
\bibitem[1970]{munstel} Mustel, E. R., Boyarchuk, A. A., 1970, ApSS, 6, 183
\bibitem[1995]{ost1} Oestreicher, M. O., Gochermann, J., Schmidt-Kaler, T., 1995, A\&AS, 112, 495
\bibitem[1996]{ost2} Oestreicher, M. O., Schmidt-Kaler, T., 1996, A\&AS, 117, 303
\bibitem[1989]{oste} Osterbrock, D. E., 1989, in {\it Astrophysics of gaseous nebulae and active galactic nuclei}, University Science Book ed.
\bibitem[1998]{sn1998} Panagia, N., 1998, MmSAI, 69, 225
\bibitem[1957]{pg} Payne-Gaposchkin, C., 1957, in {\it Book review: The Galactic Novae}, Science, 126, issue 3287, p.1350
\bibitem[1987]{pri} Pritchet, C.J., van den Bergh, S., 1987, ApJ, 318, 507
\bibitem[1991]{sch} Schwering, P. B. W., Israel, F. P., 1991, A\&A, 246, 231
\bibitem[1997]{shara} Shara, M. M., Zuerk, D. R., Williams, R.E., Prialnik, D.,  Gilmozzi, R., Moffat, A. F. J., 1997, AJ, 114, 258
\bibitem[1998]{star} Starrfield, S., Truran, J. W., Wiescher, M. C., Sparks, W. M., 1998, MNRAS, 296, 502
\bibitem[2000]{star2} Starrfield, S., Truran, J. W., Weischer, M. C., Sparks, W. M., 2000, ApJS, 127, 485
\bibitem[2002]{anu} Subramaniam, A., Anupama, G. C., 2002, A\&A, 390, 449
\bibitem[2001]{swz} Schwarz, G.J., Shore, S.N., Starrfield, S., Hauschildt, P.H., Della Valle, M., Baron, E., 2001, MNRAS, 320, 103
\bibitem[1990]{lmc902} Sekiguchi, K, Caldwell, J. A. R., Stobie, R. S., Buckley, D. A. H., 1990, MNRAS, 245, 28
\bibitem[2003]{steve}Shore, S. N., Schwarz, G., Bond, H. E., Downes, R. A., Starrfield, S., Evans, A., Gehrz, R. D., Hauschildt, P.H., Krautter, J. Woodward, C.E., 2003, AJ, 125, 1507
\bibitem[1986]{m31} van den Bergh, S., Pritchet, C.J., 1986, PASP, 98, 110
\bibitem[1987]{vanden} van den Bergh, S., Younger, P. F., 1987, A\&AS, 70, 125
\bibitem[1999]{vanlan} Vanlandingham, K. M., Starrfield, S., Shore, S. N., Sonneborn, G, 1999, MNRAS, 308, 577
\bibitem[1991]{w91} Williams , R. E., Hamuy, M., Phillips, M. M., Heathcote, S. R., Wells, L., Navarrete, M., 1991, ApJ, 376, 721
\bibitem[1992]{w92} Williams, R. E., 1992, AJ, 104, 725
\bibitem[1994]{w94s} Williams, R. E., 1994, AJ, 426, 279
\bibitem[1994]{w94} Williams, R. E., Phillips, M. M., Hamuy, M., 1994, ApJS, 90, 297
\bibitem[2004]{m33} Williams, S. J., Shafter, A. W., 2004, ApJ, 612, 867
\end{thebibliography}
\end{document}